\documentclass[aps,prl,twocolumn,showpacs,superscriptaddress]{revtex4}
\usepackage{graphicx}
\usepackage{bm}

\begin{document}

\title{Thin-Film Aluminum Microstructure as a Hot-Electron Microwave Radiation Detector}

\author{I.~Yu.~Borisenko}
\author{V.~I.~Kuznetsov}
\email[Electronic address:]{kvi@ipmt-hpm.ac.ru}
\author{V.~A.~Tulin}
\affiliation{Institute of Microelectronics Technology and High Purity Materials,
Russian Academy of Sciences, 142432 Chernogolovka, Moskow Region, Russia}
\author{D.~Esteve}
\affiliation{Quantronics group, Service de Physique de l'Etat
Condense, DSM/DRECAM, CEA Saclay, 91191 Gif-sur-Yvette, France}

\date{\today}

\begin{abstract}
We have found that the thin film aluminum structures shaped into a
chain of micron sized islands connected by narrow isthmuses, can
modify their electrical and structural properties under microwave
radiation. As a result, at the temperature of 4.2 K the film
structures turn into a kind of lateral periodic structure N-S-N,
where N is for normal islands, S is for superconducting isthmuses.
Current-voltage characteristics of the samples, as well as changes
of these characteristics under low power radiation, have been
studied over the temperature range from 1.3 to 10 K. The
sensitivity of a structure as a microwave detector runs $10^{5}$
V/W.
\end{abstract}


\maketitle

Currently, scientists apply themselves to a search and development
of highly sensitive detectors and mixers for UHF and IR diapasons.
In particular, radioastronomy makes high demands of such devices.
The up-to-date detectors are based on quantum principles. A
penetrating quantum of electromagnetic radiation (photon) produces
quasiparticles (current-carriers usually) in the material of a
detector \cite{c1,c2}. As a result, electrical properties of the
detector change. These changes are fixed by the next following
blocks of a receiver. Effective transformation of photons to
quasi-particles as well as a small or stable number of background
particles are necessary to provide a high sensitivity of a
detector. Such requirements determine the materials of the
detector, namely, they are semiconductors which possess a few
carriers or superconductors with a few normal excitations at
$T<T_{c}$, whose number is guaranteed by low working temperatures
of the detector. Contemporary detectors with a superconductor as a
sensitive element are of two types of structures:
superconductor-isolator-superconductor (SIS-detector or mixer)
\cite{c1} and superconducting submicron bridge between the sides
of a normal metal (N-S-N - hot electron detector or mixer)
\cite{c3}. Another important thing to improve the reliable work of
the detector is to match the resistance of the structure with the
impedance of the microwave system, to which the detector belongs.
The desirable magnitudes of resistance should hit into the range
from 10 to 100 $\Omega$ . To find a built-in detector of microwave
radiation, which could be suitable for low temperature devices, we
have studied the chain of metal film islands (each size was about
1 $\mu$m) connected by isthmuses 0.1 $\mu$m in width.

The main measurements were carried out on the aluminum structures.
The aluminum film about 100 nm thick was deposited on the Si
substrate masked by photoresist. To form the drawing on the
aluminum film, the liftoff process was applied. As a result, the
structure consisting of a chain of triangles (rhombuses) was
formed, All the triangles (rhombuses) were bound by the top of one
triangle to the basis of the next one (the top of the previous
rhombus to the top of the next one). The side of a triangle was 1
$\mu$m, the width of a contact was about 0.1 - 0.2 $\mu$m. Fig. 1a
presents the SEM image of the fully proceeded structure. The
structure was surrounded with the electrical gold layout, which
allowed CVC (current - voltage characteristic) to be measured.
\begin{figure}
\includegraphics [width = 1.0\linewidth]{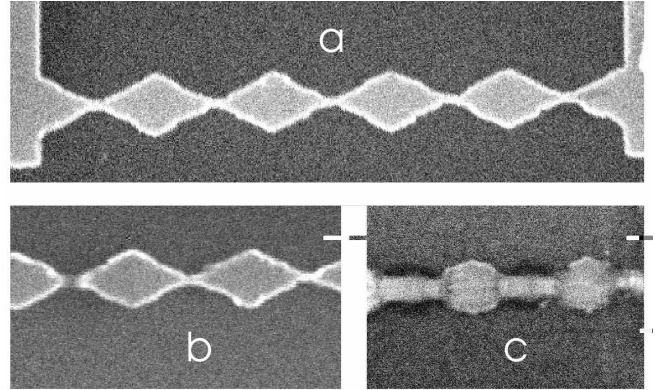}
\caption{\label{fig. 1} SEM image of the sample: a) after the
preparation process; b) after the first modification and
measurement; c) after multiple modifications.}
\end{figure}
The main experimental results of the given paper were obtained for
rhombuses. An understanding of CVC of these samples is relatively
simple as compared to CVC structures consisting of triangles.

To be irradiated by the electromagnetic field (38 GHz) the sample
was placed into the slit of the wave-guide centered against its
wide wall along its axis. The narrow wall of the wave-guide was
narrowed up to 0.5 mm in order to allow the sample to tie built
into the continuous current measuring circuit after it was placed
into the microwave cell. The narrowing of the wave-guide aimed
simultaneously to make a concentrator of the electric field. The
output power of the applied oscillator reached 10 mW. To change
the microwave field on the sample, a microwave attenuator was
applied. The prepared inset based on a wave-guide was mounted in
the helium cryostat. The setup allowed us to measure the CVC
characteristics over the temperature range from 1.3 to 10 K under
the electromagnetic irradiation of the different power. The
preliminary measurements showed the linear behaviour of CVCs of
the structures, the resistance ratio was $R_{300}/R_{4.2} \simeq
20$. The resistance of each sample at 4.2 K was about 6 $\Omega$.
Then the sample was exposed to microwave irradiation at helium
temperature (the input power at the wave-guide section containing
the sample ran up to 10 mW). The variations of CVC began to
manifest at the power of about 0.1-1 mW. Then the resistance of
the structure decreased to 2 $\Omega$ at small bias voltages and
at the same temperature of 4.2 K the samples demonstrated rather
unexpected behaviour of CVCs, which was very similar to those of
superconducting one-dimensional channels with the phase-slip
centers (Fig. 2, curve 4.2 K).
\begin{figure}
\includegraphics [width = 1.0\linewidth]{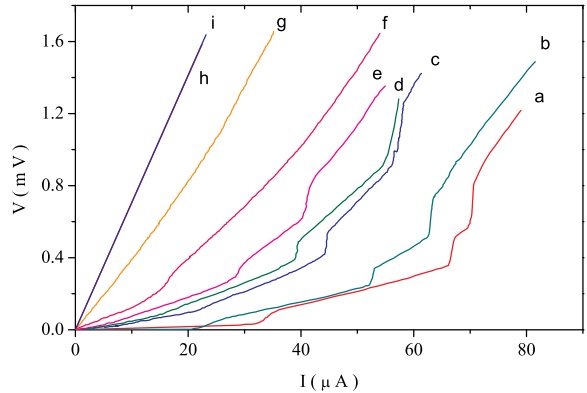}
\caption{\label{fig. 2} Current - voltage characteristics of the
sample modified by microwave irradiation at different
temperatures: (a) T=4.2 K; (b) 4.34 K; (c) 4.62 K; (d) 4.7 K; (e)
4.83 K; (f) 5 K; (g) 5.23 K; (h) 6.15 K; (i) 6.55 K.}
\end{figure}
The linear CVC corresponding to the 2 $\Omega$ resistance was
observed up to the current $I_{c1}$ (from 10 to 100 $\mu$A, for
different samples). At the current $I_{c1}$ an abrupt voltage
increase was observed, corresponding to an increase of the
resistance of the structure. As current increased the current the
CVC transformed to a broken line consisting of linear sections
connected by the sections of a voltage sharp increasing. The slope
of linear sections was divisible by 10 or 20 $\Omega$ for the
structures consisting of triangles or rhombuses, respectively.
Fig. 2 presents the CVC measured at different temperatures for one
of the samples. At temperatures higher than 6 K, the samples
demonstrate linear CVCs, which correspond to the resistance of
about 100 $\Omega$. It should be noted that about ten samples were
measured, so the above mentioned resistance are to be taken for an
order of magnitude. As the temperature decreases below 6 K, a
non-linearity occurs at small voltage shifts, which corresponds to
the abrupt decrease of the resistance in the case. The CVCs have
the form of a broken line in the temperature range from 4.2 to 1.5
K, their resistance are 2 $\Omega$ for the currents lower than
those at the first stage. The resistance disappears below 1.4 K at
this stage, and the aluminum structure transforms into the
superconducting state.

As a result of the study of more than ten samples, we concluded
that, the behaviour of the samples corresponded, on the whole, to
the properties of the periodic superconducting - normal chain.
Some parts of the chain (islands) were in the normal state while
the others (isthmuses) were superconducting. Such the unexpected
behaviour (the critical temperature of the superconducting
transition of aluminum runs 4.2 K) testifies to the change of
properties resulting from our manipulations with the sample. The
microwave irradiation can be in our opinion such a provocative
factor. It should be noted that the modification described was
obtained in all the samples under investigation. Fig. 1b presents
the SEM image of the structure which possesses the above mentioned
properties as exposed to microwave irradiation and measured in
liquid helium. Some changes can be observed nearby narrow places
as compared to Fig. 1a. They are due to the changes in structural
properties of materials (CVC changes) as well as to the material
transfer (vagueness nearby isthmuses). Fig. 1c represents the
micrograph of the sample which was multiply exposed to the
microwave irradiation. Drastic changes of the sample shape already
took place. The rhombuses were transformed to "circles", which are
connected by strips of the modified material. Noticeable migration
of the material in the sample under the microwave irradiation is
observed in an electron microscope. The structural changes also
take place, which result in new electrical properties of the
sample.

After the CVC of the sample became non-linear due to our
manipulations, we studied its variation under the low power
microwave irradiation.
\begin{figure}
\includegraphics [width = 1.0\linewidth]{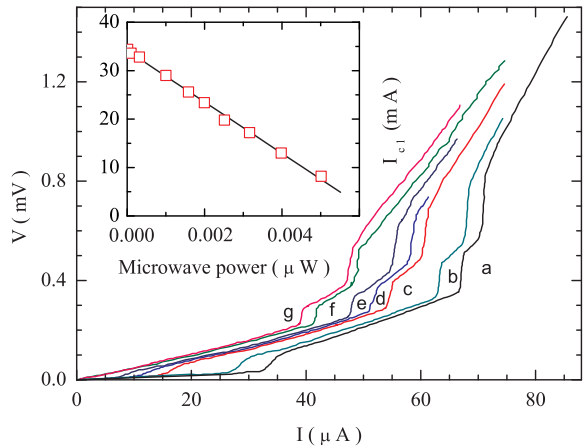}
\caption{\label{fig. 3} Change of the current - voltage
characteristic, effected by microwave irradiation of different
powers. Attenuation in decibels is specified as a parameter for
the curves: (a) 70 dB; (b) 60 dB; (c) 50 dB; (d) 45 dB; (e) 44 dB;
(f) 43 dB; (g) 41 dB. The inset shows the dependence of CVC, first
step position on the current curve, on the microwave power.}
\end{figure}
Fig. 3 presents CVC recording at 4.2 K. Damping decrement along
the microwave path is indicated in decibel as a parameter for the
curves. It is evident from Fig. 3 that microwave irradiation
results in the total voltage increment of the structure (at the
constant current) as well as in the shift of voltage increment
steps towards lower currents. The inset in Fig. 3 demonstrates the
dependence of the first CVC current step position (Fig. 3) on the
microwave radiation power at the entrance to the wave-guide
section. The position of the step is proportional to the microwave
power. The value of slope $dI_{c}/dW$ can be estimated from this
dependence. The product of this derivative by the step resistance
$dV/dI$ is the sensitivity of our structure as a detector.
According to our estimate, the value of the sensitivity is
$10^{5}$ V/W within the region of CVC steps.

Fig. 4 presents CVC of the aluminum sample (chain of rhombuses)
which was exposed to, so to speak, the minimum dose of irradiation
as follows: the current of 30 $\mu$A was conducted through the
sample at 4.2 K. The microwave oscillator was turned on by a
meander stated under the regime of modulation at the frequency of
1 kHz. The signal was transmitted from potential contacts of the
sample to an oscillograph. The attenuator of the microwave track
was gradually opened. At the particular moment (20 dB for the
given measurement) the signal of the modulated microwave power
appeared abruptly. After that the attenuator was closed and CVC
measurements were fulfilled at the larger attenuation in the
microwave track.
\begin{figure}
\includegraphics [width = 1.0\linewidth]{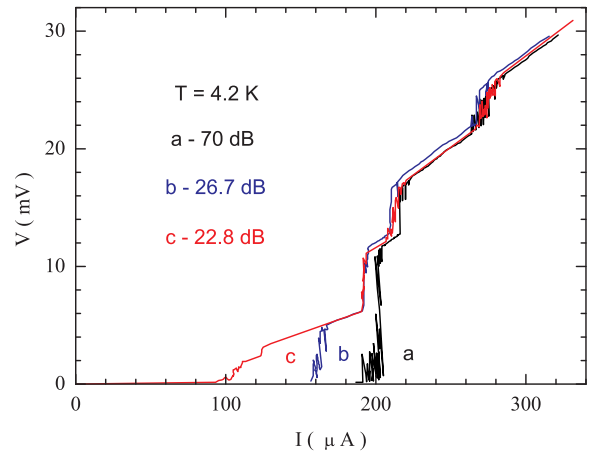}
\caption{\label{fig. 4} Current - voltage characteristic of the
sample effected by microwave irradiation. Attenuation in decibels
is specified as a parameter.}
\end{figure}
As evidenced by Fig. 4, the characteristic features of the CVC of
the sample are analogous to those presented in Fig. 3, except for
the transition between ohmic parts realized by leaps from one
state to another and vice versa. An instability manifests itself
in both superconducting and resistive states of an isthmus. It is
necessary for the existence of such an instability that the other
state of the system should be more energetically profitable than
the starting one, no matter what kind it is: resistive or
superconducting. In this case the system of carriers will try to
transform from one state to another. In the opposite case only
hysteresis would simply be observed.

As a result of manipulations at helium temperature, our periodical
aluminum structure gained new properties which are of interest
from the practical and scientific viewpoint. CVC and the SEM image
show that at 4.2 K the sample proved to be a sequential chain of
triangles (rhombuses) of the normal metal connected by
superconducting isthmuses. The initial section of CVC with the
resistance of 2 $\Omega$ corresponds to the normal state for the
general area of a triangle, except for a contact tip, while the
isthmuses proved to be superconductors with the transition
temperature up to 5.8 K. The critical current and critical
temperature are not the same for each isthmus because of their
different sizes. Consecutive abrupt voltage increments
corresponding to the transition of each isthmus to the resistive
state were observed as the resistance of all the structure
increased by the value of the isthmus resistance. These
speculations completely embrace the results of the experimental
observations of the CVCs' behaviour.

The resulting structure looks as an alternation of parts of
relatively large normal metal sections bridged by superconducting
isthmuses, which corresponds to a series of consecutive "hot
electron detectors". The structure does not change until the
temperature of the transition is reached and becomes
superconducting below this temperature (about 1.3 K). Some
advantages should be mentioned as compared to traditionally used
"hot electron detectors". The structure consists of the single
material being in two different solid states. Such the structure
can facilitate the exit of hot electrons from superconducting
parts as compared to the case of heteroboundaries in traditional
detectors, whose ability can result in the improvement of
frequency characteristics of the detector at the mode of the
diffusion exit of carriers \cite{c3}. The ohmic resistance of the
structure and its elements hit into the suitable range to provide
the matching with the microwave track or with an antenna.
Unfortunately we have no the certified devices to measure the
sensitivity at extremely low levels of the microwave power and to
compare with the background.

The behaviour of the material at the liquid helium temperature
under the action of the microwave irradiation proved to be the
most enigmatic result among the above described. Such a scaling
low-temperature migration of aluminum under ordinary conditions
has not, probably, been observed before. Material science
investigations are required in this case. First of all,  the
question arises: what is the microstructure of modified areas and
what is a role (or a mechanism) of the microwave effect? The
similar experiment carried out with the Bi samples did not show
any changes in them under the microwave irradiation. The interest
in Bi is due to the fact that the amorphous Bi modification is
superconducting at the temperature below 5.5 K.

From the viewpoint of the scientific interest the above described
samples are considered to be the lateral structure N-S-N in which
parts of films of different materials do not overlap. The
properties of isthmuses (the volume of the superconducting phase)
can relatively smoothly change directly in the process of
measurements at the helium temperature. In the light of stated
above it is interesting to carry out more detailed measurements of
CVC and their derivatives.

The experimental results obtained in the work have shown that the
aluminum samples shaped into a chain of islands connected by
narrow isthmuses modify their electric and structural properties
under the microwave irradiation at the liquid helium temperature.
As a result of the modification, they transform into the periodic
lateral structure N-S-N at 4.2 K, where N is for normal islands, S
is for superconducting isthmuses. The CVC of the samples as well
as their changes under the low power microwave irradiation were
studied at different temperatures in the range from 1.3 to 10 K.
CVCs of modified chains proved to be broken lines with the
sections of the abrupt voltage increment corresponding to the
consecutive transition of the superconducting isthmuses to the
normal metallic state. The sample sensitivity as a microwave
detector is $10^{5}$ V/W in the region of the transition between
linear sections of the CVC.

We would like to express our acknowledgments to Prof. Ye.~Glickman
for advice and for his interest in the work and to Dr. S.~Dubonos
for fine electron beam lithography.

This work was supported by the International Association for the
promotion of co-operation with scientists from the New Independent
States of the former Soviet Union (project INTAS 96-0721).

\end{document}